\documentclass[12pt]{iopart}
\makeatletter
\@namedef{ver@amsmath.sty}{}
\makeatother
\usepackage{xcolor}
\usepackage{graphicx}
\usepackage{bm}
\usepackage{amssymb}

\begin{document}

\title[Directed transport of suspended ferromagnetic nanoparticles]{Directed transport of suspended ferromagnetic nanoparticles under both gradient and uniform magnetic fields}

\author{S I Denisov, T V Lyutyy and M O Pavlyuk}
\address{Sumy State University, Rimsky-Korsakov Street 2, UA-40007
Sumy, Ukraine}
\ead{denisov@sumdu.edu.ua}

\begin{abstract}
The suspended ferromagnetic particles subjected to the gradient and uniform magnetic fields experience both the translational force generated by the field gradient and the rotational torque generated by the fields strengths. Although the uniform field does not contribute to the force, it nevertheless influences the translational motion of these particles. This occurs because the translational force depends on the direction of the particle magnetization, which in turn depends on the fields strengths. To study this influence, a minimal set of equations describing the coupled translational and rotational motions of nanosized ferromagnetic particles is introduced and solved in the low Reynolds number approximation. Trajectory analysis reveals that, depending on the initial positions of nanoparticles, there exist four regimes of their directed transport. The intervals of initial positions that correspond to different dynamical regimes are determined, their dependence on the uniform magnetic field is established, and strong impact of this field on the directed transport is demonstrated. The ability and efficiency  of the uniform magnetic field to control the separation of suspended ferromagnetic nanoparticles is also discussed.
\end{abstract}


\submitto{\JPD}

\maketitle

\section{Introduction}
\label{Intr}

Ferromagnetic single-domain nanoparticles distributed in a viscous liquid have interesting physical properties and promising biomedical applications, such as magnetic fluid hyperthermia, magnetic particle imaging, drug delivery and many others (see, e.g., recent reviews \cite{Card2018, Vall2019} and references therein). Because these applications often utilize the properties of the magnetic and mechanical dynamics of such nanoparticles, it is important to correctly introduce their corresponding equations of motion. One of the most reliable methods proposed to derive these equations is based on the concept of the total particle momentum, which includes the angular and spin momentum parts. Within this approach, the coupled equations of motion for the particle magnetization and the particle angular velocity have been derived and analyzed \cite{Usov2015, Usad2015, Usad2017, LyHrYa2019}, and a number of nontrivial effects in these systems controlled by the magnetocrystalline anisotropy have been predicted and studied \cite{Kesh2017, LyHrCo2018, LyDe2019}.

At the same time, when the anisotropy energy is large compared to other energies, the magnetization vector can be considered as `frozen' into the particle body  \cite{Shliomis1972, Coffey2004}. The main advantage of this so-called rigid dipole model is that the particle dynamics completely describes the magnetization dynamics and, as a consequence, the magnetic properties of such systems. This approximation was used, e.g., to investigate the role of the magnetic dipolar interaction and thermal fluctuations in energy dissipation \cite{PhysRevE.79.021407, PhysRevE.83.021401, Lyutyy2018a}, formation of structures due to dipolar interaction \cite{C1SM05343D, POLYAKOV20131483, PhysRevE.96.012603}, deterministic and stochastic rotation of ferromagnetic nanoparticles \cite{Raible2004, PhysRevE.92.042312, PhysRevE.97.032610} and many other phenomena.

Recently, using this model, we theoretically and numerically studied the effect of directed transport (drift) of ferromagnetic nanoparticles induced by the Magnus force \cite{DenPed2017, DENISOV201789, PhysRevE.97.032608}. It has been shown that nano\-particles performing in a viscous liquid synchronized translational oscil\-lations (induced by the oscillating driving force) and nonuniform rotations (induced by the nonuniformly rotating magnetic field) drift in a certain direction with a constant average velocity. Since the magnitude and direction of the drift velocity can be easily controlled and tuned by the external parameters, this effect could be used, e.g., in drug delivery and separation applications. But the ratio of the drift velocity to the particle velocity caused by the non-oscillating driving force is proportional to the rotational Reynolds number, which for nanosized particles is sufficiently small. As a consequence, the velocity of nanoparticles subjected to the non-oscillating driving force is usually much larger then the corresponding drift velocity.

One of the most commonly used methods to generate the driving force acting on ferromagnetic nanoparticles is the gradient magnetic field. Although the influence of this field on nanoparticles is well studied in the context of magnetic separation (see, e.g., \cite{Svob2004, Ha2013}), a complete analysis of their directed transport under the application of the gradient and uniform magnetic fields, to the best of our knowledge, has not been performed yet. Because the uniform field strongly affects the magnetization orientation, it is expected that this field can be used to control and manipulate the directed transport of suspended ferromagnetic nanoparticles induced by the gradient magnetic field. Therefore, in this paper, we aim to study the dynamics of such nanoparticles with a special focus on their transport properties and possible applications.

\section{Motion equations}
\label{Eqs}

We study the directed transport of a spherical ferromagnetic nanoparticle suspended in a viscous liquid that is charac\-terized by the radius $a$ and magnetization $\mathbf{M} = \mathbf{M}(t)$, where $|\mathbf{M}| = M = \mathrm{const}$. The particle radius is chosen to be so small that the single-domain state is realized and, at the same time, so large that the thermal fluctuations can be neglected (these conditions hold for a variety of materials, see, e.g., \cite{Guim} and below). If, in addition, the anisotropy magnetic field is strong enough, the particle can be associated with a rigid magnetic dipole. In this approximation the particle magnetization is directed along the particle easy axis and hence satisfies the kinematic differential equation
\begin{equation}
    \dot{\mathbf{M}} = \boldsymbol{
    \omega} \times \mathbf{M},
    \label{kinem}
\end{equation}
where $\boldsymbol{ \omega} = \boldsymbol{ \omega}(t)$ is the particle angular velocity, the sign $\times$ denotes the vector product, and the overdot denotes the derivative with respect to the time $t$. It is assumed also that the radius-vector of the particle center, $\mathbf{R} = \mathbf{R}(t)$, changes so slowly that the derivative $\dot{\mathbf{M}}$ can be calculated at a fixed $\mathbf{R}$.

We consider the situation when a nanoparticle is under influence of both the uniform $\mathbf{H}_{\perp}$ and gradient $\mathbf{H}_{g}$ magnetic fields:
\begin{equation}
    \mathbf{H}_{\perp} = H_{\perp}
    \mathbf{e}_{y}, \quad \mathbf{H}_{g}
    = g R_{x}\mathbf{e}_{x}.
    \label{H,H_g}
\end{equation}
Here, $H_{\perp}(\geq 0)$ is the uniform magnetic field strength, $g\,(>0)$ is the gradient of the magnetic field $\mathbf{H}_{g}$, $R_{x}$ is the $x$ component of $\mathbf{R}$, and $\mathbf{e}_{x}$, $\mathbf{e}_{y}$ and $\mathbf{e}_{z}$ are the unit vectors along the corresponding axes of the Cartesian coordinate system $xyz$. It should also be noted that the uniform magnetic field $\mathbf{H}_{\parallel} = H_{\parallel}     \mathbf{e}_{x}$ together with the gradient magnetic field $\mathbf{H}_{g}$ act as the shifted gradient magnetic field $g (R_{x} + H_{\parallel}/g) \mathbf{e}_{x}$, i.e., $\mathbf{H}_{ \parallel}$ shifts only the point where $\mathbf{H}_{g}= \mathbf{0}$. Therefore, without loss of generality, at this stage we can choose $H_{\parallel} =0$.

To describe the magnetization dynamics in these fields, we assume that the magnetization vector $\mathbf{M}$ lies in the $xy$ plane:
\begin{equation}
    \mathbf{M} = M (\cos{\varphi}
    \, \mathbf{e}_{x} +
    \sin{\varphi}\,\mathbf{e}_{y}),
    \label{def_M}
\end{equation}
where $\varphi = \varphi(t)$ is the azimuthal angle of $\mathbf{M}$. As it follows from the kinematic equation, the connection between this angle and the particle angular velocity is given by the usual relation
\begin{equation}
    \dot{\varphi} = \omega_{z}.
    \label{eq_varphi}
\end{equation}

Let us now write the equations describing the trans\-lational and rotational dynamics of a suspended particle. Because of its nanoscale size, the inertial effects can safely be neglected (see below). Therefore, keeping only the driving and friction terms, these equations can be written as the force balance equation, $\mathbf{f}_{d} + \mathbf{f}_{f} = \mathbf{0}$, and the torque balance equation, $\mathbf{t}_{d} + \mathbf{t}_{f} = \mathbf{0}$, respectively. Here, $\mathbf{f}_{d}$ is the driving force generated by the gradient magnetic field, $\mathbf{f}_{f}$ is the friction force, $\mathbf{t}_{d}$ is the driving torque exerted by the total magnetic field $\mathbf{H}_{\perp} + \mathbf{H}_{g}$, and $\mathbf{t}_{f}$ is the frictional torque. Taking into account that the  driving force and torque are defined as $\mathbf{f}_{d} = V (\mathbf{M} \cdot \partial /\partial \mathbf{R}) \mathbf{H} _{g}$ ($V = 4\pi a^{3}/3$ is the particle volume, the dot denotes the scalar product) and $\mathbf{t}_{d} = V \mathbf{M} \times (\mathbf{H} _{\perp} + \mathbf{H}_{g})$ and calculating them using (\ref{H,H_g}) and (\ref{def_M}), the above equations can be rewritten as
\begin{equation}
    MVg \cos{\varphi}\, \mathbf{e}_{x}
    + \mathbf{f}_{f} = \mathbf{0}
    \label{eq_f}
\end{equation}
and
\begin{equation}
    MV(H_{\perp} \cos{\varphi} - gR_{x}
    \sin{\varphi})\, \mathbf{e}_{z} +
    \mathbf{t}_{f} = \mathbf{0}.
    \label{eq_t}
\end{equation}
Since explicit expressions for $\mathbf{f}_{f}$ and $\mathbf{t} _{f}$ are in general not known, below we consider only the case of small Reynolds numbers (for nanoparticles this condition is not too restrictive).

\subsection{Low Reynolds number approximation}
\label{Low}

The trans\-lational and rotational Reynolds numbers defined as $\mathrm{Re} = 2\rho a|\mathbf{v}|/\eta$ and $\mathrm{Re}_{\omega} = \rho a^{2} |\boldsymbol{ \omega}|/\eta$, respectively, play the most important role in determining $\mathbf{f}_{f}$ and $\mathbf{t}_{f}$. Here, $\rho$ is the liquid density, $\mathbf{v} = \mathbf{v} (t)$ is the particle translational velocity and $\eta$ is the dynamic viscosity of liquid. If these parameters are chosen so that $\mathrm{Re} \ll 1$ and $\mathrm{Re}_{\omega} \ll 1$, then the liquid flow induced by a moving particle is laminar and, according to \cite{rubinow_1961}, $\mathbf{f}_{f} = -6\pi \eta a\mathbf{v}$ and $\mathbf{t}_{f} = -8\pi \eta a^{3} \boldsymbol{ \omega}$. Using the last formula, from  (\ref{eq_t}) we find $\boldsymbol{ \omega} = \omega_{z} \mathbf{e}_{z}$ with
\begin{equation}
    \omega_{z} = \frac{M}{6\eta} (H_{\perp}
    \cos{\varphi} - gR_{x} \sin{\varphi}).
    \label{omega_z}
\end{equation}
Then, substituting (\ref{omega_z}) into (\ref{eq_varphi}), one obtains the following equation for the azimuthal angle of the magnetization vector:
\begin{equation}
    \dot{\varphi} - \omega_{\perp}
    \cos{\varphi} + \omega_{g} r_{x}
    \sin{\varphi} = 0,
    \label{eq_varphi1}
\end{equation}
where $r_{x} = R_{x}/a$ is the dimensionless $x$ component of the particle position and
\begin{equation}
    \omega_{\perp} = \frac{MH_{\perp}}
    {6\eta}, \quad
    \omega_{g} = \frac{Mga}{6\eta}
    \label{omega_c,g}
\end{equation}
are the characteristic frequencies arising from the uniform and gradient magnetic fields, respectively. Without loss of generality, we assume that the initial azimuthal angle $\varphi(0) = \varphi_{0}$ satisfies the condition $\varphi_{0} \in [0, \pi]$.

Equation (\ref{eq_f}) with $\mathbf{f}_{f}$ given above shows that the particle velocity $\mathbf{v}$ has only the $x$ component
\begin{equation}
    v_{x} = v_{0} \cos{\varphi},
    \label{v_x}
\end{equation}
where
\begin{equation}
    v_{0} = \frac{2Mga^{2}}{9\eta}
    = \frac{4}{3}\omega_{g}a
    \label{def_v_0}
\end{equation}
is the particle characteristic velocity. Using (\ref{v_x}) and (\ref{def_v_0}), one gets
\begin{equation}
    r_{x} = r_{x0} + \frac{4}{3}\omega_{g}
    \int_{0}^{t} \cos{\varphi(t')}dt'
    \label{r_x}
\end{equation}
with $r_{x0} = r_{x}(0)$ being the initial particle position.

The set of equations (\ref{eq_varphi1}) and (\ref{r_x}) together with the initial values $\varphi_{0}$ and $r_{x0}$ completely describes the rotational and translational dynamics of such nanoparticles. According to (\ref{r_x}), equation (\ref{eq_varphi1}) is integro-differential (this is because the  strength of the gradient magnetic field acting on particles changes during their translational motion). Rewriting it in the form
\begin{equation}
    \frac{\dot{\varphi} - \omega_{\perp}
    \cos{\varphi}}{\sin{\varphi}}=
    - \omega_{g} r_{x}
    \label{eq_varphi2}
\end{equation}
and taking the time derivative of both sides, equation (\ref{eq_varphi2}) can be reduced to the autonomous second-order differential equation
\begin{equation}
    \ddot{\varphi} \sin{\varphi}-
    \dot{\varphi}^{2}\cos{\varphi}
    + \omega_{\perp} \dot{\varphi}
    + \frac{4}{3}\omega_{g}^{2}
    \sin^{2}{\varphi}\cos{\varphi}=0.
    \label{eq_dc}
\end{equation}
As (\ref{eq_varphi1}) shows, the solution of (\ref{eq_dc}) must satisfy the initial conditions $\varphi(0) = \varphi_{0}$ and
\begin{equation}
    \dot{\varphi}(0) = \omega_{\perp}
    \cos{\varphi}_{0} - \omega_{g}
    r_{x0} \sin{\varphi}_{0}.
    \label{in_cond}
\end{equation}

\section{Nanoparticle dynamics}
\label{Dyn}

Because the analysis of the nanoparticle dynamics in the cases of absence and presence of the uniform magnetic field $H_{\perp}$ is different, we consider these cases separately.

\subsection{Nanoparticle dynamics at $H_{\perp} =0$}
\label{H=0}

Since in this case $\omega_{\perp} =0$, the second-order nonlinear differential equation (\ref{eq_dc}) can be reduced to the first-order linear differential equation for $q = \dot{\varphi}^{2}$. Indeed, considering $q$ as a function of $\varphi$, i.e., $q=q(\varphi)$, and taking into account that $q'_{\varphi} = 2\ddot{\varphi}$ ($q'_{\varphi} = dq/d\varphi$), (\ref{eq_dc}) reduces to
\begin{equation}
    q'_{\varphi} \sin{\varphi} -
    2q\cos{\varphi} + \frac{8}{3}
    \omega_{g}^{2}\sin^{2}{\varphi}
    \cos{\varphi} =0
    \label{eq_q}
\end{equation}
According to (\ref{in_cond}), the solution of this equation must satisfy the condition $q(\varphi_{0}) = q_{0}$, where
\begin{equation}
    q_{0} = \omega_{g}^{2}r_{x0}^{2}
    \sin^{2}{\varphi}_{0}.
    \label{q_0}
\end{equation}

The general solution of (\ref{eq_q}) is given as follows (see, e.g., formula (13.1.4) in \cite{PoZa2018}):
\begin{equation}
    q = C\sin^{2}{\varphi} - \frac{4}{3}
    \omega_{g}^{2} \sin^{2}{\varphi}\ln{
    \sin^{2}{\varphi}},
    \label{sol_q1}
\end{equation}
where $C$ is the integration constant. Determining this constant from the condition (\ref{q_0}),
\begin{equation}
    C = \omega_{g}^{2}r_{x0}^{2}
    + \frac{4}{3} \omega_{g}^{2}
    \ln{\sin^{2}{\varphi}_{0}},
    \label{C}
\end{equation}
and using (\ref{def_v_0}), the solution (\ref{sol_q1}) can be rewritten in the form
\begin{equation}
    q = \omega_{g}^{2} \,\Bigg(
    r_{x0}^{2} - \frac{4}{3}
    \ln{\frac{\sin^{2}{\varphi}}
    {\sin^{2}{\varphi}_{0}}}\Bigg)
    \sin^{2}{\varphi}.
    \label{sol_q2}
\end{equation}
From this and equation (\ref{eq_varphi1}), which at $H_{\perp} =0$ yields $q = \dot{\varphi}^{2} = \omega_{g}^{2} r_{x}^{2} \sin^{2}{\varphi}$, one obtains the relation
\begin{equation}
    r_{x}^{2} = r_{x0}^{2} -\frac{4}
    {3} \ln{\frac{\sin^{2}{\varphi}}
    {\sin^{2}{\varphi}_{0}}}
    \label{r_x^2}
\end{equation}
that plays an important role in the further analysis of the nano\-particle dynamics.

We begin our analysis by noting that the initial angular velocity $\dot{\varphi}(0) = - \omega_{g} r_{x0} \sin{ \varphi}_{0}$ depends on the initial particle position $r_{x0}$. This fact, together with (\ref{eq_varphi1}), which according to (\ref{r_x^2}) can be represented as
\begin{equation}
    \dot{\varphi} = -\mathrm{sgn}\,
    (r_{x})\,\omega_{g}\, \Bigg(
    r_{x0}^{2} - \frac{4}{3} \ln{
    \frac{\sin^{2}{\varphi}}{\sin^{2}
    {\varphi}_{0}}}\Bigg)^{1/2}
    \sin{\varphi}
    \label{eq_varphi4}
\end{equation}
[$\mathrm{sgn}\,(x) = \pm 1$ if $x\gtrless 0$], indicates that the linear particle velocity $v_{x} = v_{0} \cos{\varphi}$ at $t \neq 0$ also depends on $r_{x0}$ (recall in this connection that the initial velocity $v_{x}(0) = v_{0} \cos{\varphi_{0}}$ is the same for all $r_{x0}$). Assuming for definiteness that $\varphi_{0} \in (0, \pi/2)$, from (\ref{eq_varphi4}) it follows that the azimuthal angle $\varphi$ monotonically decreases with time from $\varphi_{0}$ to 0 (and so $v_{x}$ monotonically increases from $v_{0} \cos{\varphi_{0}}$ to $v_{0}$) if $r_{x0} \geq 0$. In other words, all particles with $r_{x0} \geq 0$ move to the right with velocities approaching $v_{0}$ at long times. Moreover, the larger the initial particle position is, the faster the particle velocity approaches the limiting value $v_{0}$. For the purpose of classification, we call the nanoparticle dynamics at $r_{x0} \geq 0$ as the first dynamical regime.

The second dynamical regime occurs for nanoparticles with $r_{x0} \in (-l, 0)$, where
\begin{equation}
    l= \Bigg(\frac{4}{3} \ln{
    \frac{1}{\sin^{2}{\varphi}_{0}}}
    \Bigg)^{1/2}.
    \label{def_l}
\end{equation}
In this case all particles also move to the right. However, in contrast to the previous case, the azimuthal angle $\varphi$ initially increases from $\varphi_{0}$ to some $\varphi_{m} < \pi/2$ (until the position $r_{x} = 0$ is reached) and then monotonically decreases to $0$ as $t \to \infty$. In accordance with this, the particle velocity initially decreases from $v_{0}\cos{\varphi_{0}}$ to $v_{0}\cos{\varphi_{m}}$ and then grows to $v_{0}$.

The nanoparticle dynamics at $r_{x0} = -l$ corresponds to the third dynamical regime, which can be considered as a limiting case of the second one. As before, the particle initially moves to the right, but after reaching the state with $r_{x}=0$ and $\varphi_{m} = \pi/2$ its motion is stopped. We note, however, that this state is unstable: due to fluctuations, the particle leaves the vicinity of this point and moves either to the left or to the right.

Finally, the fourth dynamical regime is realized if $r_{x0} < -l$. In this case the azimuthal angle $\varphi$ monotonically increases with time from $\varphi_{0}$ to $\pi$, and each particle moves to the right only on the time interval $(0, t_{s})$, which depends on $r_{x0}$. At $t=t_{s}$ the particle stops [$v_{x}(t_{s}) =0$, i.e., $\varphi(t_{s}) = \pi/2$] at the point $r_{x}(t_{s}) = -\big(r_{x0}^2 - l^2 \big)^{1/2}$, and then (at $t>t_{s}$) moves to the left reaching the velocity $-v_{0}$ in the long-time limit. As for $r_{x0} >0$, the larger $|r_{x0}|$ is, the faster the limiting velocity $-v_{0}$ is reached.

It should be also mentioned that in the special case when $\varphi_{0} = \pi/2$ only two dynamical regimes, the first and fourth, can be realized at $r_{x0} >0$ and $r_{x0} <0$, respectively. Since in this case $l=0$, the second and third regimes are reduced to the state $r_{x} = r_{x0} =0$, which is unstable (the particle with $r_{x0}=0$ moves either to the left or to the right).

To illustrate the theoretical and numerical results, we consider $\mathrm{SmCo}_{5}$ nanoparticles suspended in water at room temperature ($295\, \mathrm{K}$) and characterized by the parameters $M = 1.36 \times 10^{3}\, \mathrm{emu\, cm^{-3}}$, $\rho_{n} = 8.31\, \mathrm{g\, cm^{-3}}$ is the particle density, $\rho = 1\, \mathrm{g\, cm^{-3}}$, and $\eta = 9.62\times 10^{-3}\, \mathrm{P}$. Choosing $a = 2\times10^{-5}\, \mathrm{cm}$ (the critical single-domain diameter for these particles is about $7.5\times 10^{-5}\, \mathrm{cm}$ \cite{Goya2008}) and $g = 10^{2}\, \mathrm{Oe\, cm^{-1}}$, one finds $v_{0} = 1.26 \times 10^{-3}\, \mathrm{cm\, s^{-1}}$ and $\omega_{g} = 47.12\, \mathrm{s^{-1}}$. With these parameters, replacing $|\mathbf{v}|$ by $v_{0}$ and $|\boldsymbol{ \omega}|$ by $MgR_{x}/6\eta$ with $|R_{x}| = 1\, \mathrm{cm}$, the definitions of $\mathrm{Re}$ and $\mathrm{Re}_{\omega}$ yield $\mathrm{Re} = 5.24\times 10^{-6}$ and $\mathrm{Re}_{\omega} = 9.8\times 10^{-2}$. Since the representation $\mathbf{t}_{f} = -8\pi \eta a^{3} \boldsymbol{ \omega}$ holds even for $\mathrm{Re}_{\omega} \lesssim 10$ \cite{Dennis1980}, the approximation of small Reynolds numbers is well justified. In our model we also neglect the inertial terms $\rho_{n} V\dot{\mathbf{v}}$ and $J\dot{\boldsymbol{ \omega}}$ [$J = (2/5)\rho_{n} Va^{2}$ is the particle moment of inertia] in equations (\ref{eq_f}) and (\ref{eq_t}), respectively. A simple analysis shows that these terms can indeed be neglected at $t \gg \mathrm{max}\{t_{tr}, t_{r} \}$, where
\begin{equation}
    t_{tr} = \frac{2\rho_{n}a^{2}}
    {9\eta},
    \quad
    t_{r} = \frac{\rho_{n}a^{2}}
    {15\eta}
    \label{t_rel}
\end{equation}
are the translational and rotational relaxation times. Because, according to (\ref{t_rel}), $\mathrm{max}\{t_{tr}, t_{r} \} = t_{tr} = 7.68\times 10^{-8}\, \mathrm{s}$, we make sure that the inertial effects in the dynamics of $\mathrm{ SmCo}_{5}$ nanoparticles are negligible already at $t \gg 10^{-7}\, \mathrm{s}$. Finally, the single-particle approximation used in our theoretical model is justified, i.e., the magnetic dipole-dipole and hydrodynamic interactions can be ignored, if the average distance $d$ between nanoparticles is large enough. In particular, the energy of the dipole-dipole interaction of two particles, $(MV)^{2}/d^{3}$, is negligible compared to the particle energy in the gradient magnetic field, $MV|\mathbf{H} _{g}|$, if $d \gg (4\pi M/ 3g|R_{x}|)^{1/3}a$. At the same time, the condition of smallness of the hydrodynamic interaction, which holds when the volume fraction of nanoparticles is small, i.e., $V/d^{3} \ll 1$, is not so restrictive: $d \gg (4\pi/3)^{1/3}a$.

\begin{figure}[ht]
    \centering
    \includegraphics[totalheight=5.5cm]{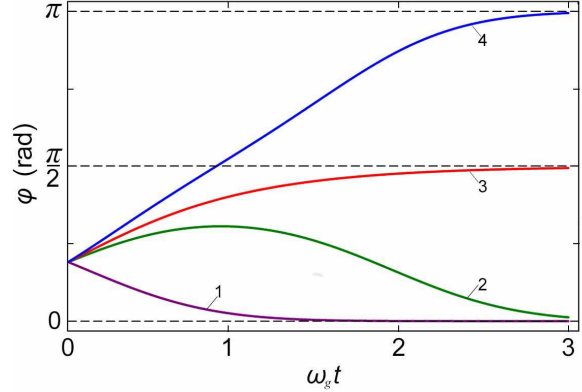}
    \caption{Plots of the function $\varphi =
    \varphi(t)$ obtained via numerical solution
     of equation (\ref{eq_dc}) for $H_{\perp}=0$,
     $\varphi_0 = 0.6 \, \mathrm{rad}$ and different
     values of the initial particle position $r_{x0}$.
     Since, according to (\ref{def_l}), in this case
     $l = 1.235$, the values of $r_{x0}$ are
     chosen to reproduce all predicted regimes
     of the nanoparticle dynamics: $r_{x0} = 1.0$
     (curve 1), $r_{x0} = -1.0$ (curve 2), $r_{x0}
     = -l$ (curve 3), and $r_{x0} = -1.5$ (curve 4).}
    \label{fig1}
\end{figure}

\begin{figure}[ht]
    \centering
    \includegraphics[totalheight=5.5cm]{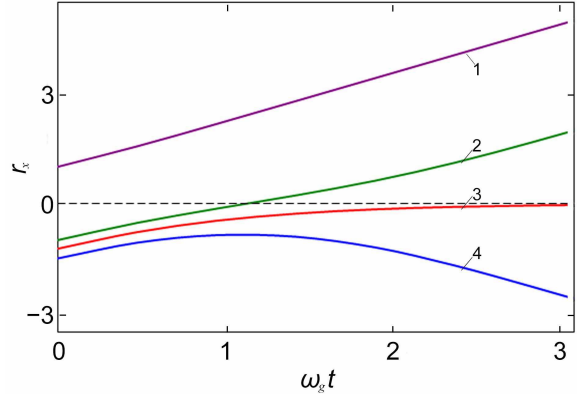}
    \caption{Plots of the dimensionless particle
    position $r_{x} = r_{x}(t)$ obtained from
    (\ref{r_x}) by solving equation (\ref{eq_dc}) for
    the same parameters as in figure \ref{fig1}.}
    \label{fig2}
\end{figure}

\begin{figure}[ht]
    \centering
    \includegraphics[totalheight=5.5cm]{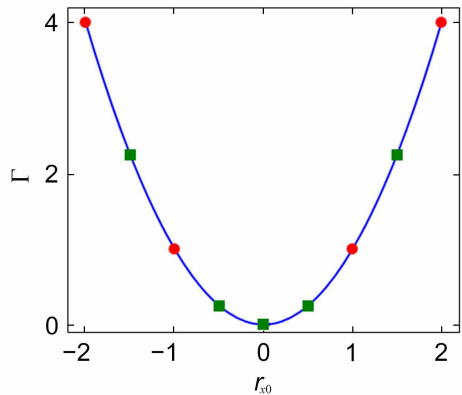}
    \caption{Dependence of the quantity
    $\Gamma$ on the dimensionless initial
    particle position $r_{x0}$. The numerical
    data are marked by symbols (the circles and
    squares correspond to $\omega_{g}t= 1$
    and $\omega_{g}t= 3$, respectively), and
    the theoretical result $\Gamma =r_{x0}^{2}$
    is shown by the solid curve.}
    \label{fig3}
\end{figure}

In figure \ref{fig1}, we show the dependence of the azimuthal angle $\varphi$ on the dimensionless time $\omega_{g} t$ for different values of $r_{x0}$.
They represent four regimes of behavior of the function $\varphi=\varphi(t)$, when $r_{x0} \geq 0$ (first regime, curve 1), $r_{x0} \in (-l,0)$ (second regime, curve 2), $r_{x0} = -l$ (third regime, curve 3), and $r_{x0} < -l$ (fourth regime, curve 4). Figure \ref{fig2} illustrates the time dependence of the dimensionless particle coordinate $r_{x}$ under the same conditions as in figure \ref{fig1}.
As seen, the numerical results presented in these figures confirm the existence of predicted regimes of the nanoparticle dynamics. Finally, to verify the theoretical result (\ref{r_x^2}), we used the numerical results from figures \ref{fig1} and \ref{fig2} to calculate the quantity $\Gamma = r_{x}^{2} + (4/3) \ln{(\sin^{2} \varphi /\sin^{2}\varphi_{0})}$ for two moments of time and different values of the initial particle position $r_{x0}$. The comparison of the obtained data with the theoretical dependence $\Gamma = r_{x0}^{2}$, see figure \ref{fig3}, confirms its validity.

As it was mentioned above, the magnetic field $\mathbf{H}_{ \parallel}$ shifts the point in which $\mathbf{H}_{g} = \mathbf{0}$. This means that $\mathbf{H}_{ \parallel}$ shifts also the intervals where different regimes of the nanoparticle dynamics are realized (for example, the first dynamical regime occurs now at $r_{x0} > - H_{ \parallel}/ga$). It is important to emphasize that, since the dimensionless shift value $|H_{ \parallel}|/ga$ can be rather large even for small $H_{ \parallel}$ (e.g., in our case it equals $10^{4}$ if $H_{ \parallel} = 20\, \mathrm{Oe}$), the varying of the magnetic field strength $H_{ \parallel}$ is an effective method for changing the dynamical regimes.

\subsection{Nanoparticle dynamics at $H_{\perp} \neq 0$}
\label{H_neq_0}

In this case, the theoretical analysis of transport properties of ferromagnetic nano\-particles is more complicated. Therefore, here we study them analytically only at small and large times. In order to analyse the nanoparticle dynamics at $t \to 0$, it is convenient to use equations (\ref{eq_varphi1}) and (\ref{r_x}), which in the first-order approximation yield
\begin{equation}
    \varphi = \varphi_{0} + \big(\omega_{
    \perp} \cos{\varphi}_{0} - \omega_{g}
    r_{x0} \sin{\varphi}_{0} \big)\, t
    \label{varphi1}
\end{equation}
and
\begin{equation}
    r_{x} = r_{x0} + \frac{v_{0}}{a}
    \cos{\varphi_{0}}\,t.
    \label{r_x2}
\end{equation}
The last result shows that, as in the previous case, all particles at small times move to the right with the same initial velocity $v_{0} \cos{\varphi_{0}}$ [recall, $\varphi_{0} \in (0, \pi/2)$]. In contrast, the rotation of particles depends now not only on $r_{x0}$, but also on $H_{\perp}$.

Our qualitative analysis suggests that, like for $H_{\perp} = 0$, four regimes of the nanoparticle dynamics exist for $H_{\perp} \neq 0$ as well. They are realized at $r_{x0} \geq l_{1}$, $r_{x0} \in (-l_{2}, l_{1})$, $r_{x0} = -l_{2}$, and $r_{x0} < -l_{2}$, respectively. According to (\ref{varphi1}), the azimuthal angle $\varphi$ decreases monotonically with time, i.e., the first dynamical regime is realized, if $r_{x0} \geq l_{1}$, where
\begin{equation}
    l_{1} = \frac{\omega_{\perp}}
    {\omega_{g}} \cot{\varphi_{0}}.
    \label{def_l1}
\end{equation}
This result agrees with that for $H_{\perp} = 0$ ($l_{1} \to 0$ as $H_{\perp} \to 0$) and indicates that the magnetic field $H_{\perp}$ can significantly change the value of $l_{1}$ (the condition $l_{1} \gg 1$ can easily be achieved).

Because our analysis is approximate, we are not able to find an exact expression for $l_{2}$. However, using (\ref{varphi1}) and (\ref{r_x2}), it is possible to roughly estimate this quantity. Indeed, choosing $r_{x0} = -l_{2}$ and assuming that $\varphi{(t_{s})} = \pi/2$ and $r_{x}(t_{s}) = 0$ (recall, $t_{s}$ is the stoping time), from (\ref{varphi1}) -- (\ref{def_l1}) one obtains
\begin{equation}
    l_{2} = \frac{1}{2} l_{1} \Bigg[
    \Bigg( 1 + \frac{8(\pi - 2\varphi_{0})}
    {3l_{1}^{2}\tan{\varphi_{0}}}
    \Bigg)^{1/2} -1 \Bigg].
    \label{def_l2}
\end{equation}
For $l_{1} \ll 1$ ($H_{\perp} \ll ga\tan{\varphi_{0}}$) and $l_{1} \gg 1$ ($H_{\perp} \gg ga\tan{\varphi_{0}}$) this formula yields
\begin{equation}
    l_{2}|_{l_{1} \ll 1} =
    \Bigg( \frac{2(\pi - 2\varphi_{0})}
    {3\tan{\varphi_{0}}}
    \Bigg)^{1/2}
    \label{l2_1}
\end{equation}
and
\begin{equation}
    l_{2}|_{l_{1} \gg 1} =
    \frac{2}{3}(\pi - 2\varphi_{0})
    \,\frac{ga}{H_{\perp}}
    \label{l2_2}
\end{equation}
($l_{2}|_{l_{1} \gg 1} \ll 1$). We emphasize two points regarding these results. First, the approximate formula (\ref{l2_1}) is in perfect agreement with the exact one (\ref{def_l}) only if $\pi/2 - \varphi_{0} \ll 1$; the difference between $l_{2}|_{l_{1} \ll 1}$ and $l$ grows rapidly with decreasing $\varphi_{0}$. This is a consequence of the fact that the solutions (\ref{varphi1}) and (\ref{r_x2}) are valid only for $\omega_{g}t \ll 1$. And second, since the condition $l_{1} \gg 1$ is easily achieved, the magnetic field $H_{\perp}$ can also be used for changing the transport properties of suspended nanoparticles. It should be especially noted that $H_{\perp}$, in contrast to $H_{\parallel}$, shifts the regions of different dynamical regimes nonuniformly.

At large times, it is convenient to introduce the parameter $\sigma$, which for particles moving to the right or left equals 1 or $-1$, respectively. Since the azimuthal angle $\varphi$ for such particles tends to zero or $\pi$, it can be represented as $\varphi = \pi (1-\sigma)/2 + \sigma \varphi_{1}$ with $\varphi_{1} \ll 1$. Taking also into account that, in linear approximation, $\cos{\varphi} = \sigma$, $\sin{\varphi} = \varphi_{1}$ and, according to (\ref{r_x}), $r_{x} \sim \sigma (4/3) \omega_{g}t$ as $t \to \infty$, equation (\ref{eq_varphi1}) in the long-time limit reduces to
\begin{equation}
    \dot{\varphi}_{1} - \omega_{\perp}
    + \frac{4}{3}\omega_{g}^{2} t\,
    \varphi_{1} = 0.
    \label{eq_varphi_1}
\end{equation}
Its asymptotic solution is given by $\varphi_{1} \sim 3\omega_{\perp} /(4\omega_{g}^{2} t)$, i.e., the azimuthal angle $\varphi$ approaches the limiting values 0 and $\pi$ inversely proportional to time. If $\omega_{\perp} =0$, these limiting values are approached exponentially: $\varphi_{1} \sim \exp{(-2 \omega_{g}^{2} t^{2}/3)}$.

\begin{figure}[ht]
    \centering
    \includegraphics[totalheight=5.5cm]{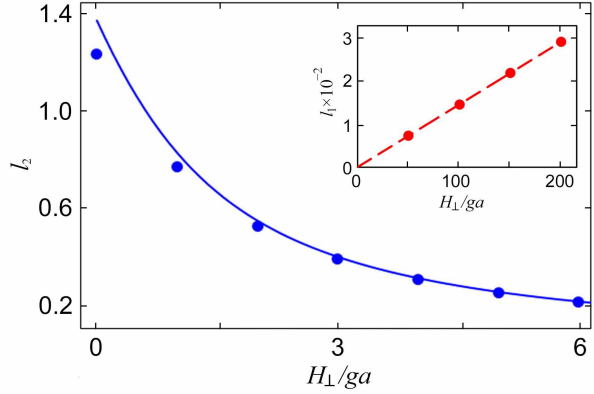}
    \caption{Dependence of $l_{2}$ and $l_{1}$
    (in the inset) on the dimensionless magnetic
    field $H_{\perp}/ga$ for the same parameters
    as in figure \ref{fig1}. The numerical data
    for $l_{1}$ and $l_{2}$, obtained by solving
    equation (\ref{eq_dc}), are marked by circles.
    The approximate formula (\ref{def_l2}) and
    theoretical result (\ref{def_l1}) are represented
    by solid and dashed lines, respectively.}
    \label{fig4}
\end{figure}

Numerical analysis of the nanoparticle dynamics at $H_{\perp} \neq 0$ confirms both the existence of four dynamical regimes (similar to those for $H_{\perp}=0$) and strong influence of $H_{\perp}$ on the intervals of $r_{x0}$, where these regimes occur. In figure \ref{fig4}, we show the dependence of the boundaries of these intervals, $l_{1}$ and $l_{2}$, on the ratio $H_{\perp}/ga$ ($=\omega_{\perp} / \omega_{g}$). Since, according to (\ref{l2_2}), $l_{2}$ approaches zero as $H_{\perp}$ increases, the width $\Delta = l_{1} + l_{2}$ of the interval $(-l_{2}, l_{1})$, where the second dynamical regime is realized, is of the order of $l_{1} (\gg 1)$ even for small $H_{\perp}$ (e.g., $\Delta \approx l_{1} = 731$ for $H_{\perp} = 1\, \mathrm{Oe}$). Recall in this context that $l_{1}|_{H_{\perp} =0} =0$, $l_{2}|_{H_{\perp} =0} =l$ and so $\Delta|_{H_{\perp} =0} = 1.235$.  As seen from this figure, the theoretical and numerical results for $l_{1}$ are in complete agreement. Note also that in spite of the approximate character of the theoretical result (\ref{def_l2}), it agrees perfectly with the numerical data for $l_{2}$.
\begin{figure}[ht]
    \centering
    \includegraphics[totalheight=5.5cm]{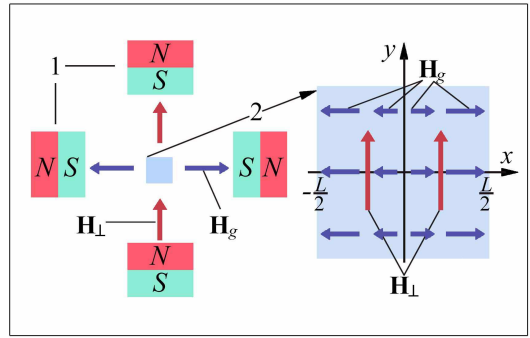}
    \caption{Schematic diagram of the system
    and sources for the uniform and gradient
    magnetic fields. The electromagnets
    generating the uniform ($\mathbf{H}_{\perp}$)
    and gradient ($\mathbf{H}_{g}$) magnetic
    fields and the suspension reservoir are
    marked by 1 and 2, respectively.}
    \label{fig5}
\end{figure}

Thus, the dynamics of suspended ferromagnetic nanoparticles in the gradient magnetic field is strongly affected by uniform magnetic fields $H_{\parallel}$ and $H_{\perp}$. It is important to emphasize that while $H_{\parallel}$ only shifts the zero point of the gradient field, the influence of $H_{\perp}$ on the nanoparticle dynamics is much more complicated. It seems that the non-trivial transport properties of these particles, resulting from the joint action of the gradient magnetic field and uniform magnetic fields $H_{\parallel}$ and $H_{\perp}$, could be used, e.g., in separation science.

In order to illustrate the feasibility and effectiveness of the separation process induced by the gradient and uniform magnetic fields, let us estimate the time $T$ of complete separation of suspended $\mathrm{SmCo}_{5}$ nanoparticles (whose parameters are given above) in a rectangular reservoir, see figure \ref{fig5}. For this purpose, we assume that the reservoir boundaries are perpendicular to the coordinate axes, the $x$ axis crosses the left and right boundaries at the points $x = -L/2$ and $x= L/2$, respectively ($L$ is the reservoir width in the $x$ direction), and the number of suspended nanoparticles in the reservoir equals $N$. In this geometry, after switching on the gradient magnetic field, the nanoparticles are concentrated near the left and right boundaries. Assuming also that the distribution of nanoparticles at $t=0$ is uniform and the limiting velocities $v_{0}$ and $-v_{0}$ are achieved for $t \ll T$, one may conclude that if $H_{\parallel} = H_{\perp} =0$, then the complete separation with $N_{l} \approx N/2$ and $N_{r} \approx N/2$ ($N_{l}$ and $N_{r}$ are the number of nanoparticles near the left and right boundaries, respectively), is achieved at $T \approx L/2v_{0}$ (e.g., $T \approx 6.6\, \mathrm{min}$ for $L=1\, \mathrm{cm}$). At the same time, if $H_{\parallel} = gL/2$ and $H_{\perp} =0$, then $N_{l} \approx 0$ and $N_{r} \approx N$ at $T \approx L/v_{0} \approx 13.2\, \mathrm{min}$. It is important to stress that real suspensions contain nanoparticles of different sizes. Because the characteristic velocity $v_{0}$ depends on the particle size, see (\ref{def_v_0}), the separation time $T$ is different for different nanoparticles. In particular, for smaller $\mathrm{SmCo}_{5}$ nanoparticles with $a = 10^{-5}\, \mathrm{cm}$ we have $v_{0} = 3.15 \times 10^{-4}\, \mathrm{cm\, s^{-1}}$, and so $T \approx 26.4\, \mathrm{min}$ if $H_{\parallel} = H_{\perp} =0$ and $T \approx 52.8\, \mathrm{min}$ if $H_{\parallel} = gL/2$ and $H_{\perp} =0$, respectively, i.e., the smaller the nanoparticles the slower the separation process. Note also that the concentration profile of nanoparticles for $t \in (0, T)$ and arbitrary $H_{\parallel}$ and $H_{\perp}$ can easily be calculated within the above theory.

\section{Conclusions}
\label{Concl}

The gradient magnetic field produces the force that acts on suspended ferromagnetic nanoparticles and induces their translational motion along the gradient field direction. In contrast, since the external uniform magnetic field does not produce any force, this field does not affect directly the translational motion of these particles. However, if particles are subjected to both the gradient and uniform magnetic fields, the latter can influence their transport properties. The reason is that the force caused by the magnetic field gradient depends on the direction of the particle magnetization. Therefore, changing the magnetization direction, the uniform magnetic field (as well as the gradient magnetic field) can indirectly affect the transport properties of suspended ferromagnetic nanoparticles.

To study this effect in detail, we have introduced a minimal set of equations that describes the coupled translational and rotational dynamics of suspended nanoparticles under the action of the gradient and uniform magnetic fields. By solving these equations analytically and numerically, we have surprisingly discovered that nanoparticles exhibit complex dynamical behavior. In particular, it has been established that, depending on the initial particle positions, there exist four different regimes for the directed transport of such nanoparticles. Namely, the particle velocity in these regimes (I) increases with time and then saturates, (II) decreases and then increases to the saturated value, (III) decreases to zero, and (IV) decreases to zero, changes sign and saturates again. It has also been shown that the external uniform magnetic field significantly changes the intervals of the initial particle positions, where these regimes are realized. Based on these properties, we have proposed to use the gradient and uniform magnetic fields for controllable separation of suspended nanoparticles. It seems also that the observed properties of directed transport may be useful for such biomedical applications as drug delivery and cell separation \cite{Ul2016, Da2019}.

\ack{This work was partially supported by the Ministry of Edu\-ca\-tion and Science of Ukraine under Grant No.\ 0119U100772.}


\section*{References}
\bibliographystyle{unsrt}
\bibliography{DirectedTransport}


\end{document}